\documentclass[11pt,showpacs,preprintnumbers,superscriptaddress,amsmath,amssymb,nofootinbib]{revtex4}
\usepackage{graphicx}
\usepackage{dcolumn}
\usepackage{bm}
\usepackage{amssymb}
\usepackage{amsmath}
\usepackage{epsfig}    
\usepackage{color}
\usepackage{slashed}
\usepackage{float}
\newcolumntype{I}{!{\vrule width 1.3pt}}

\def\be{\begin{equation}}
\def\ee{\end{equation}}
\newcommand{\bea}{\begin{eqnarray}}
\newcommand{\eea}{\end{eqnarray}}
\newcommand{\nn}{\nonumber}

\def\dsl#1{#1\hspace{-2.1mm}\slash}

\begin{document}
\title{Renormalizable Model for Neutrino Mass, Dark Matter, Muon $g-2$\\ and 750 GeV Diphoton Excess}
\author{Hiroshi Okada}
\affiliation{School of Physics, KIAS, Seoul 130-722, Korea}
\affiliation{Physics Division, National Center for Theoretical Sciences, Hsinchu, Taiwan 300}
\author{Kei Yagyu}
\affiliation{School of Physics and Astronomy, University of Southampton, Southampton, SO17 1BJ, United Kingdom}

\begin{abstract}

We discuss a possibility to explain the 750 GeV diphoton excess 
observed at the LHC in a three-loop neutrino mass model which has 
a similar structure to the model by Krauss, Nasri and Trodden. 
Tiny neutrino masses are naturally generated by the loop effect 
of new particles with their couplings and masses to be of order 0.1-1 and TeV, respectively. 
The lightest right-handed neutrino, which runs in the three-loop diagram, can be a dark matter candidate. 
In addition, the deviation in the measured value of the muon anomalous magnetic moment from its prediction in the standard model
can be compensated by one-loop diagrams with exotic multi-charged leptons and scalar bosons. 
For the diphoton event, an additional isospin singlet real scalar field plays the role to explain the excess by taking its mass of 750 GeV, where
it is produced from the gluon fusion production via the mixing with the standard model like Higgs boson. 
We find that the cross section of the diphoton process can be obtained to be a few fb level by taking the masses of new charged particles to be about 375 GeV
and related coupling constants to be order 1.

\end{abstract}

\maketitle

\section{Introduction}

In December 2015, the both ATLAS and CMS groups have reported 
the existence of the excess at around 750 GeV in the diphoton distribution at the Large Hadron Collider (LHC) with the collision energy of 13 TeV. 
The local significance of this excess is 
about 3.6$\sigma$ at ATLAS~\cite{750GeV-ATLAS} 
with the integrated luminosity of 3.2 fb$^{-1}$ and about 2.6$\sigma$ at CMS~\cite{750GeV-CMS}  
with the integrated luminosity of 2.6 fb$^{-1}$. 
The detailed properties of the diphoton excess was summarized, e.g., in Ref.~\cite{Pomarol}, where 
the best fit value of the width of the new resonance is about 45 GeV, and the estimated cross section of the diphoton signature 
is $10\pm 3$ fb at ATLAS and $6\pm 3$ fb at CMS. 
If this excess is confirmed by future data, it suggests the existence of a new particle which gives the direct evidence of a new physics beyond the standard model (SM). 

The simplest way to explain this excess is to consider an extension of the SM by adding 
extra isospin scalar multiplets such as a singlet, a doublet  and/or a triplet. 
However, it is difficult to get a sufficient cross section to explain the excess as mentioned in the above 
in such a simple extension of the SM. 
For example, if we consider the CP-conserving two Higgs doublet models (THDMs)~\cite{Moretti,thdm0,thdm-Scalar,thdm-VLF,thdm-SUSY}, and take  
the masses of the additional CP-even $H$ and CP-odd $A$ Higgs bosons to be 750 GeV, then the cross section of $pp\to H/A\to \gamma\gamma$
is typically three order smaller than the required value~\cite{Moretti}. 
Therefore, we need to further introduce additional sources to get an enhancement of the production cross section and/or the branching fraction 
to the diphoton mode, e.g., by introducing multi-charged scalar particles~\cite{Moretti,thdm-Scalar} and vector-like fermions~\cite{thdm-VLF}.  
In Refs.~\cite{thdm-SUSY}, the diphoton excess has been discussed in supersymmetric models. 

In this paper, we discuss a scenario to {\it naturally} introduce multi-charged particles to get an enhancement of the branching fraction. 
Namely, we consider a radiative neutrino mass model in which multi-charged particles play a role not only to increase the branching fraction but also 
to explain the smallness of neutrino masses and the anomaly of the muon anomalous magnetic moment. 
A dark matter (DM) candidate can also successfully be involved as a part of the model.  
There are a few papers discussing the diphoton excess within radiative neutrino mass models~\cite{750-rad}. 
In particular, we discuss a new three loop neutrino mass model\footnote{Other variations of three loop neutrino mass models have also been proposed in Refs.~\cite{AKS,GNR,Kajiyama,Culjak,Okada-Yagyu,Nishiwaki}. } whose structure is similar to the model by 
Krauss, Nasri and Trodden in 2003~\cite{KNT}, 
because the three loop suppression factor $1/(16\pi^2)^3$ is a suitable amount to reproduce 
the measured neutrino masses, i.e., ${\cal O}(0.1)$ eV, by order 0.1-1 couplings and TeV scale masses of new particles. 
In our model, an additional isospin real singlet scalar field can explain the diphoton excess, where it is produced from the gluon fusion process through the 
mixing with the SM-like Higgs boson. 
 
The plan of the paper is as follows. In Sec.~II, we define our model, and give the Lagrangian for the lepton sector and the scalar potential. 
In Sec.~III, we discuss the neutrino masses, the phenomenology of DM including the relic abundance and 
direct search experiments, and new contributions to the muon $g-2$. 
The diphoton excess is discussed in Sec.~IV. 
Our conclusion is summarized in Sec.~V.

\begin{center}
\begin{table*}[t]
{\small
\hfill{}
\begin{tabular}{c||c|c|c|c|c||c|c|c|c}\hline\hline 
&\multicolumn{5}{c||}{Lepton Fields}&\multicolumn{4}{c}{Scalar Fields}  \\\hline 
& $L_L^i$ & $ e_R^i $ &  $L_{5/2}^a=(L^{--a},L^{---a})^T$ &  $ E^{--a} $ &$N_R^a$ & $\Phi$ & $\kappa^{++}$  & $S^{++}$ & $\Sigma$ \\\hline
$(SU(2)_L,U(1)_Y)$ & $(\bm{2},-1/2)$ & $(\bm{1},-1)$ & $(\bm{2},-5/2)$ & $(\bm{1},-2)$ & $(\bm{1},0)$
& $(\bm{2},1/2)$   & $(\bm{1},2)$  & $(\bm{1},2)$ & $(\bm{1},0)$   \\\hline
$Z_2$  & $+$ & $+$ & $-$ & $-$ & $-$ & $+$ & $+$  & $-$ & $+$   \\\hline\hline
\end{tabular}}
\hfill{}
\caption{Particle contents and charge assignments under $SU(2)_L\times U(1)_Y\times Z_2$. 
The superscripts $i$ and $a$ denote the flavor of the SM fermions and the exotic fermions with $i=1\text{-}3$  and $a=1\text{-}N_E$, respectively. }

\label{tab:1}
\end{table*}
\end{center}

\section{The Model}

\begin{figure}[t]
\begin{center}
\includegraphics[scale=0.7]{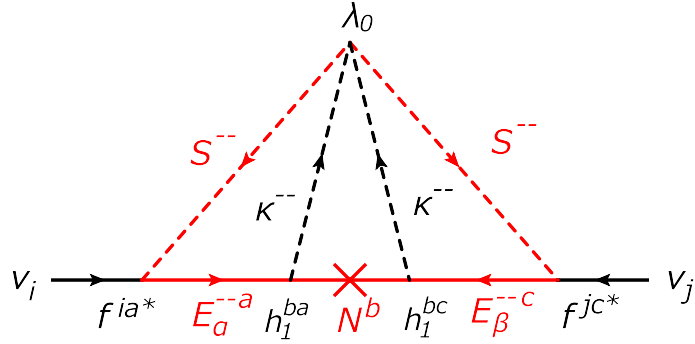}
   \caption{Three-loop neutrino mass diagram. Particles indicated by the red color are $Z_2$ odd, where 
$E_\alpha^{--a}$ and $E_\beta^{--c}$ denote the mass eigenstates for the doubly-charged exotic leptons. 
The subscripts $\alpha$ and $\beta$ (run over 1 and 2) label two mass eigenstates for each flavor. 
 }
   \label{diag}
\end{center}
\end{figure}

Our model is described by the SM gauge symmetry $SU(2)_L\times U(1)_Y$ and an additional discrete symmetry $Z_2$ which is assumed to be unbroken.  
This $Z_2$ symmetry  is introduced to avoid tree level contributions to neutrino masses and to enclose the three-loop diagram as shown in Fig.~\ref{diag}. 
Because of the $Z_2$ symmetry, the stability of the lightest neutral $Z_2$ odd particle is guaranteed, and thus it can be a candidate of DM. 

The particle contents are shown in Table~\ref{tab:1}, where 
$L_L^i$ and $e_R^{i}$ are the SM left-handed lepton doublets and lepton singlets with the flavor of $i$ ($i=$1-3). 
In addition, we add the $N_E$ flavor of the vector like lepton doublets (singlets) $L_{5/2}^a=(L^{--a},L^{---a})^T$ ($E^{--a}$) with the hypercharge $Y=-5/2$ ($-2$) and 
the right-handed neutrinos $N_R^a$ ($a=1$-$N_E$). 
The scalar sector is composed of one isospin doublet field $\Phi$ with $Y=1/2$ and 
two complex (one real) isospin singlet scalar fields $\kappa^{++}$ and $S^{++}$ with $Y=2$ ($\Sigma$ with $Y=0$). 
The doublet and the real singlet scalar fields are parameterized by 
\begin{align}
\Phi =\begin{pmatrix}
G^+\\
\frac{v+\phi^0+iG^0}{\sqrt{2}}
\end{pmatrix}, \quad \Sigma  = \sigma^0 + v_\sigma^{}, 
\end{align}
where $v$ and $v_\sigma^{}$ are the vacuum expectation values (VEVs) 
of doublet and singlet scalar fields, respectively, and $G^+$ $(G^0)$ denotes the Nambu-Goldstone boson
which is absorbed into the longitudinal component of the $W$ $(Z)$ boson. 
The Fermi constant $G_F$ is given by the usual relation, i.e., $G_F = 1/(\sqrt{2}v^2)$ with $v\simeq 246$ GeV. 
The singlet VEV $v_\sigma^{}$ does not contribute to the electroweak symmetry breaking. 
We note that the shift $v_\sigma^{}\to v_\sigma'$ does not change any physical quantities, because its impact can be absorbed by the redefinition of the parameters 
in the Lagrangian. 
We thus take $v_\sigma^{}=0$ in the following discussion to make some expressions to be a simple form. 

The most general Lagrangian for the lepton fields is given by 
\begin{align}
-\mathcal{L}_{\text{lep}}
&= \frac{1}{2}M_N^a \overline{N^{ac}_R} N_R^a + M_L^a \overline{(L_{5/2}^a)_L} (L_{5/2}^a)_R + M_E^a \overline{E^{--a}_L} E^{--a}_R + {\rm h.c.} \nn \\
& +y_{\text{SM}}^{i}\, \overline{L_L^i} \Phi\, e_{R}^i 
 +y_1^{ab}\, \overline{(L_{5/2}^a)_L} \tilde{\Phi}\, E_R^{--b} + y_2^{ab}\, \overline{(L_{5/2}^a)_R} \tilde{\Phi}\, E_L^{--b} + {\rm h.c.} \notag\\
& +g_L^{ab}\, \overline{(L_{5/2}^a)_L} (L_{5/2}^a)_R\, \Sigma + g_E^{ab}\, \overline{(E^{--a})_L} \, E_R^{--b}\, \Sigma  + g_N^{ab}\overline{N_R^{ac}}\,N_R^b\,\Sigma + {\rm h.c.}\notag\\
& + h_0^{ij}\, \overline{e_R^{ic}}\,e_R^j \, \kappa^{++} + h_1^{ab}\,\overline{N_R^a}E^{--b}_L\, \kappa^{++}
  + h_2^{ab}\,\overline{N_R^{ac}}E^{--b}_R \kappa^{++} + f^{ia}\,\overline{L^i_L}(L_{5/2}^a)_R\, S^{++} + {\rm h.c.}, 
\end{align}
where $\tilde{\Phi}= i\sigma_2\Phi^*$. 
We can take the diagonal form of the invariant masses 
$M_N^a$, $M_L^a$ and $M_E^a$ for the vector like leptons $L_{5/2}^a$, $E^a$ and right-handed neutrinos $N_R^a$, respectively, without loss of generality.  
The SM leptons $L_L$ and $e_R^{}$ are taken to be the mass eigenstates, so that the Yukawa coupling $y_{\text{SM}}^i$ is given by the diagonal form. 
For simplicity, we assume that all the above parameters are real.

The most general Higgs potential is given by 
\begin{align}
&V(\Phi,\kappa^{++},S^{++},\Sigma) = V_{\text{HSM}}(\Phi,\Sigma) + 
  \mu_{\kappa}^2(\kappa^{++}\kappa^{--}) + \mu_{S}^2(S^{++}S^{--})  \nn\\
&+ A_{ \Sigma \kappa}\, \Sigma\,(\kappa^{++}\kappa^{--}) + A_{\Sigma S} \Sigma\,(S^{++}S^{--}) \notag\\
&+ \lambda_{\Phi \kappa} (\Phi^\dagger\Phi)(\kappa^{++}\kappa^{--}) + \lambda_{\Phi S} (\Phi^\dagger\Phi)(S^{++}S^{--})
+ \lambda_{\Sigma \kappa}^{} \Sigma^2 (\kappa^{++}\kappa^{--}) + \lambda_{\Sigma S}^{} \Sigma^2(S^{++}S^{--})\notag\\
&+\lambda_\kappa (\kappa^{++}\kappa^{--})^2 + \lambda_S(S^{++}S^{--})^2 
+ \lambda_{\kappa S}(\kappa^{++}\kappa^{--})(S^{++}S^{--}) \notag\\
&+[\lambda_{0}(\kappa^{++}S^{--})(\kappa^{++}S^{--}) + \text{h.c.}],  \label{pot}
\end{align}
where the complex phase of the $\lambda_0$ parameter can be absorbed by rephasing the scalar fields. 
The squared masses of the doubly-charged scalar bosons $S^{\pm\pm}$ and $\kappa^{\pm\pm}$ are given by 
\begin{align}
m_{\kappa^{\pm\pm}}^2 = \mu_{\kappa}^2 + \frac{v^2}{2}\lambda_{\Phi \kappa}, \quad
m_{S^{\pm\pm}}^2 = \mu_{S}^2 + \frac{v^2}{2}\lambda_{\Phi S}. 
 \end{align}
In Eq.~(\ref{pot}), the $V_{\text{HSM}}$ part is given as the same form as in the Higgs singlet model (HSM) involving $\Phi$ and $\Sigma$ as
\begin{align}
V_{\text{HSM}}(\Phi,\Sigma) =&\mu_\Phi^2(\Phi^\dagger\Phi)+\lambda (\Phi^\dagger\Phi)^2 
+A_{\Phi \Sigma}(\Phi^\dagger\Phi) \Sigma+ \lambda_{\Phi \Sigma} (\Phi^\dagger\Phi) \Sigma^2\notag\\
&+t_\Sigma^{}\Sigma +\mu^2_{\Sigma}\Sigma^2+ A_\Sigma\Sigma^3+ \lambda_\Sigma\Sigma^4.  \label{Eq:HSM_pot}
\end{align} 
Two CP-even scalar states $\phi^0$ from the doublet and $s^0$ from the singlet 
are mixed with each other via the mixing angle $\alpha$ defined as 
\begin{align}
\begin{pmatrix}
\sigma^0 \\ \phi^0
\end{pmatrix}=
\begin{pmatrix}
\cos\alpha & -\sin\alpha \\ 
\sin\alpha & \cos\alpha
\end{pmatrix}
\begin{pmatrix}
H \\ h
\end{pmatrix}.  \label{alpha}
\end{align}
We define $h$ as the SM-like Higgs boson with the mass of about 125 GeV which is identified as the discovered Higgs boson at the LHC.  
The detailed expressions for the masses of the CP-even Higgs bosons and the mixing angle $\alpha$ 
in terms of the potential parameters are 
given, e.g., in Ref.~\cite{HSM}. 

The masses of the exotic charged leptons are obtained from two sources, i.e., the invariant mass terms $M_E$ and $M_L$ 
and the Yukawa interaction terms $y_1$ and $y_2$.
The mass of the triply-charged leptons $L^{---a}$ is simply given by $M_L^a$. 
For the doubly-charged leptons, there is a mixing between $L^{--a}$ and $E^{--a}$ through the $y_1$ and $y_2$ terms. 
The mass matrix is given assuming $y_{1}^{ab} = y_{2}^{ab} = y_E^{a}=\text{diag}(y_E^1,\dots, y_E^{N_E^{}})$ by 
\begin{align}
{\cal L}_{\text{mass}} &= -(\overline{E^{--a}},\overline{L^{--a}})
\begin{pmatrix}
M_E^a &   M_{D}^a\\
M_{D}^a & M_L^a
\end{pmatrix}
\begin{pmatrix}
E^{--a} \\
L^{--a}
\end{pmatrix} 
=
-(\overline{E_1^{--a}},\overline{E_2^{--a}})
\begin{pmatrix}
M_{E_1}^a &   0 \\
0  & M_{E_2}^a
\end{pmatrix}
\begin{pmatrix}
E_1^{--a} \\
E_2^{--a}
\end{pmatrix}, 
\end{align}
where $M_{D}^a =  \frac{v}{\sqrt{2}}y_E^a$. 
The mass eigenstates $E_1^a$ and $E_2^a$ are defined by the orthogonal transformation:
\begin{align}
\begin{pmatrix}
E^{--a} \\
L^{--a}
\end{pmatrix}
=
\begin{pmatrix}
\cos\theta_a & -\sin\theta_a \\
\sin\theta_a & \cos\theta_a
\end{pmatrix}
\begin{pmatrix}
E_1^{--a} \\
E_2^{--a}
\end{pmatrix}. 
\end{align}
The mass eigenvalues ($M_{E_1}^a\leq M_{E_2}^a$) and the mixing angles $\theta_a$ are given by 
\begin{align}
M_{E_{1,2}}^a &= \frac{1}{2}\left(M_E^a + M_L^a \mp\sqrt{(M_E^a - M_L^a)^2 + 4(M_{D}^a)^2}\right), \\
\tan2\theta_a & = \frac{2M_D^a}{M_E^a - M_L^a}. 
\end{align}

\section{Neutrino Mass, Dark Matter, Muon $g-2$}

\subsection{Neutrino Mass}

The leading contribution to the active neutrino mass matrix $m_\nu$ is given at three-loop level as shown 
in Fig.~\ref{diag}. 
One- and two-loop diagrams which have been systematically classified in Refs.~\cite{loop1,loop2} are absent in our setup. 
The three-loop diagram is computed as follows
\begin{align}
&(m_{\nu})_{ij} = \frac{\lambda_0}{(16\pi^2)^3 M^2_{\rm max}}\notag\\
&\times \sum_{a,b,c=1}^3\sum_{\alpha,\beta=1}^2c_{\alpha\beta}
f^{ia*}\,\sin2\theta_a  M^a_{E_\alpha} h^{ba}_1 M_N^b h_1^{bc}  \, \sin2\theta_c\,M^c_{E_\beta} f^{jc*} 
F(r_{E_\alpha^a}^{}, r_{N^b}, r_{E_\beta^c},r_{S^{\pm\pm}},r_{\kappa^{\pm\pm}}),
\label{mnu1}
\end{align}
where we 
define $r_X^{} \equiv (m_X^{}/M_{\rm max})^2$ 
with $M_{\rm max}={\rm Max}(M_{E_\alpha^a},M_{E_{\beta}^c}, M_{N^b},m_{S^{\pm\pm}}^{},m_{\kappa^{\pm\pm}})$ and $m_X^{}$ is the mass of a particle $X$, 
and $c_{\alpha\beta}=1\,(-1)$ for $\alpha=\beta\,(\alpha \neq \beta)$. 
The three loop function $F$ is given by 
\begin{align}
&F(r_{E_\alpha^a}^{}, r_{N^b}, r_{E_\beta^c},r_{S^{\pm\pm}},r_{\kappa^{\pm\pm}}) =
 \int_0^1dx_1dy_1dz_1 \delta(1-x_1-y_1-z_1)\frac{1}{z_1^2-z_1}\notag\\
&\times \int_0^1dx_2dy_2dz_2 \delta(1-x_2-y_2-z_2)\frac{1}{z_2^2-z_2}\,
 \int_0^1dx_3dy_3dz_3 \delta(1-x_3-y_3-z_3)\frac{1}{\Delta_3},  \label{int}
\end{align}
where 
\begin{align}
&\Delta_3 = x_3 r_{E_\alpha^a}^{} + y_3 r_{S^{\pm\pm}} + z_3\Delta_2, \\
& \text{with}~~\Delta_2 = -\frac{x_2 r_{N^b}^{} + y_2 \Delta_1 + z_2 r_{\kappa^{\pm\pm}}}{z_2^2-z_2},~~ 
\text{and}~~\Delta_1 = -\frac{xr_{E_\beta^c}^{} + yr_{S^{\pm\pm}} +z r_{\kappa^{\pm\pm}}}{z_1^2-z_1}. 
\end{align}
The interval of the integrals in Eq.~(\ref{int}) for all the variables is from 0 to 1, i.e., $\int_0^1 dxdydz = \int_0^1 dx\int_0^1dy\int_0^1dz$. 
Typical values of $F$ with $r_i=(0.1,1)$ are ${\cal O} (1)$. 
Let us estimate magnitudes of couplings and masses to reproduce the magnitude of neutrino masses, i.e., the  order of 0.1 eV. 
For simplicity, when we take $M_{\text{max}} = M_{E_\alpha^a}\sim M_{E_\beta^c} \sim M_{N^b} $, 
the neutrino masses are approximately expressed as 
\begin{align}
(m_{\nu})_{ij} &\sim \frac{M_{\rm max}}{(16\pi^2)^3} \times K_{ij}\sim
(0.1~\text{eV})\times 10^{6} \times \left(\frac{M_{\text{max}}}{v}\right)\times K_{ij}, \\
&\text{with}~~K_{ij} = \sum_{a,b,c=1}^3  f^{ia*}\, h^{ba}_1 h_1^{bc}   f^{jc*}, 
\end{align}
where we assume $\lambda_0 \times F = {\cal O}(1)$. 
Therefore, in the range of $M_{\text{max}}=v\text{-}10\,v$, 
the magnitude of the mixing factor $K_{ij}$ is required to be ${\cal O}(10^{-7}\text{-}10^{-6})$.

\subsection{Dark Matter}

Assuming that the right-handed neutrino $N_R^1$ is the lightest among all the $Z_2$ odd particles, 
$N_R^1$ looses its decay modes into any other lighter particles, and then it becomes stable. 
We thus can regard $N_R^1$ as the DM candidate in our model. 
The annihilation cross section is then calculated as 
\begin{align}
&\sigma v_{\rm rel}\approx 
\int_0^{\pi} d\theta\sin\theta \int_0^{2\pi}d\phi\frac{1}{128\pi^2 s} |\overline{\cal M}(N_R^1N_R^1 \to AB)|^2 \sqrt{1-\frac{4 m^2_{\text {fin}  }}{s}},  \label{ann}
\end{align}
where $m_{\text{fin}}$ is the mass of the final state particle. 
In the above expression, $|\overline{\cal M}(N_R^1N_R^1 \to AB)|^2$ is the squared amplitude for the following two body to two body processes:
\begin{align}
&|\overline{\cal M}(N_R^1N_R^1 \to AB)|^2  = |\overline{\cal M} (N_{R}^1N_{R}^1  \to \kappa^{++}\kappa^{--})|^2+
|\overline{\mathcal{M}}(N_{R}^1N_{R}^1 \to f\bar{f})|^2 + |\overline{\mathcal{M}}(N_{R}^1N_{R}^1 \to ZZ)|^2\nn\\
&+ |\overline{\mathcal{M}}(N_{R}^1N_{R}^1 \to W^+W^-)|^2
+|\overline{\mathcal{M}}(N_{R}^1N_{R}^1 \to hh)|^2
+ |\overline{\mathcal{M}}(N_{R}^1N_{R}^1 \to HH)|^2,
\end{align}
The first annihilation process $N_R^1N_R^1  \to \kappa^{++}\kappa^{--}$ happens through 
the $t$- and $u$-channels of the $E_\alpha^a$ mediation, where 
the doubly-charged scalar bosons $\kappa^{\pm\pm}$ decays into the same sign dilepton  via the Yukawa coupling $h_0$. 
The squared amplitude of the $N_R^1N_R^1  \to \kappa^{++}\kappa^{--}$ process is given by
\begin{align}
&|\overline{\cal M} (N_{R}^1N_{R}^1  \to \kappa^{++}\kappa^{--})|^2=\sum_{a=1}^{N_E}
|h^{1a}|^2{\rm tr}\left[ (\dsl p_2- M_N^1) X_a (\dsl p_1+M_N^1) X_a^\dag \right],\\
&X_a = \cos^2\theta_a\left[\frac{-\dsl p_1+\dsl k_1+M^a_{E_1} }{t-(M^a_{E_1})^2 }+\frac{-\dsl p_1+\dsl k_2+M^a_{E_1} }{u-(M^a_{E_1})^2 }\right]
+(\cos\theta_a \to \sin\theta_a,\, M^a_{E_1} \to M^a_{E_2}),
\end{align}
where $s$, $t$ and $u$ 
are the Mandelstam variables, $N_{c}^f$ is the color factor, 
and $(p_1, p_2)$ and $(k_1, k_2)$ are the initial and the final state momenta, respectively.
In this expression, we take $h^{ab}\equiv h_1^{ab}=h_2^{ab}$ for simplicity.  
The other cross sections are given through the mixing of $\alpha$ via the $s$-channel mediation of $h$ and $H$ by
\begin{align}
|\overline{\mathcal{M}}(N_R^1N_R^1 \to f\bar{f})|^2
&=
16 {N_c^f} \left( \frac{g^{11}_N m_f s_{\alpha} c_{\alpha}}{v} \right)^2
 \left| \frac{1}{s - m_h^2 + im_h\Gamma_h} - \frac{1}{s - m_H^2 + im_H \Gamma_H} \right|^2 \notag \\
&\quad \times\left[ (p_1 \cdot p_2) - (M_N^1)^2 \right] \left[ (k_1 \cdot k_2) - m_f^2 \right], \\
|\overline{\mathcal{M}}(N_{R}^1N_{R}^1 \to ZZ)|^2
&=
8 \left( \frac{g^{11}_N m_Z^2 s_{\alpha} c_{\alpha}}{v} \right)^2
\left| \frac{1}{s - m_h^2 + im_h\Gamma_h} - \frac{1}{s - m_H^2 + im_H\Gamma_H} \right|^2  \notag \\
&\quad \times \left[ (p_1 \cdot p_2) - (M_{N}^1)^2 \right] \left[ 2 + \frac{(k_1 \cdot k_2)^2}{m_Z^4} \right], \\
|\overline{\mathcal{M}}(N_{R}^1N_{R}^1 \to W^+W^-)|^2
&=
16 \left( \frac{g^{11}_N  m_W^2 s_{\alpha} c_{\alpha}}{v} \right)^2
\left| \frac{1}{s - m_h^2 + im_h\Gamma_h} - \frac{1}{s - m_H^2 + im_H\Gamma_H} \right|^2  \notag \\
&\quad \times \left[ (p_1 \cdot p_2) - (M_{N}^1)^2 \right] \left[ 2 + \frac{(k_1 \cdot k_2)^2}{m_W^4} \right], \\
|\overline{\mathcal{M}}(N_{R}^1N_{R}^1 \to hh)|^2
&=
2 \left( {g^{11}_N }{} \right)^2
\left| \frac{s_\alpha \lambda_{hhh}}{s - m_h^2 + im_h\Gamma_h} - \frac{c_\alpha \lambda_{Hhh} }{s - m_H^2 + im_H\Gamma_H} \right|^2  \notag \\
&\quad \times\left[ (p_1 \cdot p_2) - (M_{N}^1)^2 \right], \\
|\overline{\mathcal{M}}(N_{R}^1N_{R}^1 \to HH)|^2
&=
2 \left( {g^{11}_N}{} \right)^2
\left| \frac{s_\alpha \lambda_{hHH}}{s - m_h^2 + im_h\Gamma_h} - \frac{c_\alpha \lambda_{HHH} }{s - m_H^2 + im_H\Gamma_H} \right|^2  \notag \\
&\quad \times \left[ (p_1 \cdot p_2) - (M_N^1)^2 \right],
\end{align}
where we use the short-hand notations of $c_\alpha \equiv \cos{\alpha}$ and $s_{\alpha} \equiv \sin{\alpha}$.
The dimensionful $\lambda_{\varphi_i\varphi_j\varphi_k}$ couplings ($\varphi_{i,j,k}=h$ or $H$) are defined by the coefficient of the scalar trilinear vertex in the potential. 
We note that the $s$-wave contribution to $\sigma v_{\rm rel}$ vanishes due to the Majorana property of the DM.
To reproduce the observed relic density, 
the cross section given in Eq.~(\ref{ann}) should be inside the following region
\begin{align}
\sigma{v_{\rm rel}}=(1.78\text{-}1.97)\times10^{-9}\ {\rm GeV^{-2}},
\end{align}
at the $2\sigma$ level~\cite{Ade:2013zuv}. 

We also consider the spin independent scattering cross section with a neutron that is induced via the tree level diagram with the Higgs boson $h$ and $H$ exchange.
The formula is given by
\begin{align}
\sigma_{\text{SI}}^{n} =
\frac{C^2}{\pi}  \left(\frac{m_n^2 M_N^1}{m_n + M_{N}^1 }\right)^2   \left( \frac{ g^{11}_N c_{\alpha} s_{\alpha}}{v} \right)^2
\left( -\frac{1}{m_h^2} + \frac{1}{m_H^2} \right)^2,
\end{align}
where the neutron mass is $m_n\simeq 0.939$ GeV and the factor $C\simeq 0.287^2$ is determined by the lattice simulation.
The latest upper bound is reported by the LUX experiment that suggests $\sigma_{\text{SI}}^{n}\lesssim 10^{-45}$ cm 
for the DM mass of about 100 GeV with the 90 \% C.L.~\cite{LUX}.

\subsection{Muon $g-2$}

The muon anomalous magnetic moment (muon $g-2$) is one of the most promising low energy observables which 
suggest the existence of new physics beyond the SM. 
This is because there is the more than 3$\sigma$ deviation in the SM prediction from 
the experimental value measured at Brookhaven National Laboratory.  
The difference $\Delta a_{\mu}\equiv a^{\rm exp}_{\mu}-a^{\rm SM}_{\mu}$ has been calculated in Ref.~\cite{discrepancy1} as 
\begin{align}
\Delta a_{\mu}=(29.0 \pm 9.0)\times 10^{-10}. 
\end{align}
This shows the $3.2\sigma$ deviation in the SM prediction. 

In our model, two diagrams contribute to $\Delta a_\mu$, where 
$L^{---a}$-$S^{--a}$ and $\ell^{-}$-$\kappa^{--a}$ with $\ell^-$ being the SM lepton are running in the loop. 
These contributions are calculated by  
\begin{align}
\Delta a_\mu \simeq \frac{m_\mu^2}{16\pi^2}\Bigg\{&
\sum_{a=1}^{N_E}|f^{\mu a}|^2\left[\frac{3}{M_L^{a2}}G\left(\frac{m_{S^{\pm\pm}}^2}{M_L^{a2}}\right)+\frac{2}{m_{S^{\pm\pm}}^2}G\left(\frac{M_L^{a2}}{m_{S^{\pm\pm}}^2}\right)   \right] 
-\sum_{i=1}^{3}|\bar{h}_0^{\mu i}|^2\frac{2}{3m_{\kappa^{\pm\pm}}^2}\Bigg\}
\end{align}
where $\bar{h}_0^{ij}=h_0^{ij}\, (2h_0^{ij})$ for $i=j\,(i\neq j)$, and 
\begin{align}
G(x) = \frac{1-6x+3x^2+2x^3-6x^2\ln x}{(1-x)^4}. 
\end{align} 
We can see that the contribution from the $\kappa^{\pm\pm}$ loop gives the negative value which is undesired to explain the muon $g-2$ anomaly. 
We thus neglect the $\kappa^{\pm\pm}$ loop contribution that can be realized by taking $\bar{h}_0^{\mu i} \ll f^{\mu a}$.

\subsection{A set of solution}

Here, we show a set of the solution to give the sizable amount of $\Delta a_\mu$, i.e., 
$2.0\times 10^{-9}\lesssim\Delta a_{\mu}\lesssim 3.8\times 10^{-9}$,  
the non-relativistic cross section to satisfy the observed relic density $\sigma{v_{\rm rel}}=(1.78\text{-}1.97)\times10^{-9}\ {\rm GeV^{-2}}$, 
and to satisfy the constraint of the direct detection $\sigma_{\text{SI}}^{n}\lesssim 10^{-45}$, 
where we conservatively take the constraint of the direct detection for all the mass region of DM. 
By taking the number of the flavor $N_E=3$, we find the following benchmark parameter sets:
\begin{align}
&m_{\kappa^{\pm\pm}}=375\ {\rm GeV},~ m_{S^{\pm\pm}}=377\ {\rm GeV},~ M_{E_1} = 375\ {\rm GeV},~M_{E_2} = 380\ {\rm GeV},\nn\\
& 
M_{N}^1 =478\ {\rm GeV},~ M_{N}^{2,3}=556\ {\rm GeV}, ~
m_H=750\ {\rm GeV},~\Gamma_H=2.40\ {\rm GeV},   \notag\\
& \sum_{a=1}^3|f^{\mu a}|^2=3.04^2,~ \sum_{a=1}^3|h^{1a}|^2 = 0.721^2,~ g_N^{11}=1.06\times 10^{-3}, 
~ \sin\theta_E=0.141, ~ \sin\alpha = -0.1, 
\end{align}
where $M_{E_{1,2}}=M_{E_{1,2}}^a$, $\theta_E = \theta_a$ for $a = 1$-3, and we take $\lambda_{hhh} = \lambda_{Hhh}= 0$. 
The triply-charged lepton mass $M_L$ is given about 375 GeV from the above inputs. 
The values for three parameters $m_H$, $\sin\alpha$ and $\Gamma_H$ are favored 
for the discussion of the 750 GeV diphoton signature which will be discussed in the next section. 
The other SM parameters are fixed as follows  
\begin{align}
&\Gamma_h=4.1\ {\rm MeV},~ m_h=125.5\ {\rm GeV},~ v=246\ {\rm GeV},~ m_W = 80.4\ {\rm GeV},~ m_Z = 91.2\ {\rm GeV}, \notag\\
&m_t = 173\ {\rm GeV},~m_b = 4.18\ {\rm GeV}. 
\end{align}
From the benchmark set, we obtain the following results
\begin{align}
\Delta a_\mu=3.54\times10^{-9},~  \sigma{v_{\rm rel}}= 1.87\times 10^{-9}\ {\rm GeV^{-2}}.\label{eq:sol}
\end{align}

\section{Diphoton excess}

We discuss how we can reproduce the diphoton excess at around 750 GeV at the LHC. 
In our model, the additional CP-even Higgs boson $H$ plays the role to explain this excess via the gluon fusion production process by taking its mass of 750 GeV. 
The cross section $\sigma_{\gamma\gamma}$ of the diphoton channel is expressed by using the narrow width approximation as follows
\begin{align}
\sigma_{\gamma\gamma} \equiv \sigma(gg \to H \to \gamma\gamma)
 \simeq \sigma(gg \to H) \times {\cal B}(H\to \gamma\gamma). 
\end{align}
Non-zero production cross section $\sigma(gg \to H)$ of the gluon fusion process is 
given through the mixing $\alpha$ with the SM-like Higgs boson $h$ defined in Eq.~(\ref{alpha}) as 
\begin{align}
\sigma(gg \to H) = \sin^2\alpha \times \sigma(gg \to h_\text{SM}), 
\end{align}
where $h_\text{SM}$ denotes the SM Higgs boson, and $\sigma(gg \to h_\text{SM})$ does its gluon fusion cross section in which 
the mass of $h_{\text{SM}}$ here is fixed to be 750 GeV in order to derive the cross section for $H$. 
From~\cite{SM-cross}, we obtain $\sigma(gg \to h_\text{SM})\simeq$ 736 fb at the collision energy of 13 TeV. 

Next, we discuss the decays of $H$ and $h$ to figure out the branching fraction of ${\cal B}(H\to \gamma\gamma)$ and the signal strength $\mu_{XY}^{}$
of $pp \to h \to XY$ modes for $h$. 
The latter quantity becomes important to set a constraint on the parameter space.  
In particular, when we consider the enhancement of ${\cal B}(H\to \gamma\gamma)$, this could also significantly modify the 
event rates of $h$ for various channels. 
The definition of $\mu_{XY}^{}$ is given by 
\begin{align}
\mu_{XY}^{} = \sigma(gg \to h) \times {\cal B}(h \to XY). 
\end{align}

The decay rates of ${\cal H} \to {\cal PP'}$ with ${\cal H}=h$ or $H$ 
and ${\cal PP'}= f\bar{f},~W^+W^-,~ZZ$ or $gg$ are given by 
\begin{align}
\Gamma({\cal H} \to {\cal PP'}) = \xi_{\cal H}^2 \Gamma(h_\text{SM} \to {\cal PP'}), 
\end{align}
where $\xi_{\cal H}= \sin\alpha\,(\cos\alpha)$ for ${\cal H}=H\,(h)$. 
For the $\gamma\gamma$ and $Z\gamma$ modes, the decay rate is not simply given by the above way due to the additional loop contributions of the new charged particles. 
In order to simplify the discussion, we take flavor universal valuables for 
the masses of the exotic charged leptons and the mixing angles, i.e., $M_{E_\alpha}^a = M_{E_\alpha}$ and $\theta_a = \theta_E$ as we have done it in the previous section. 
In this case, 
the decay rates for ${\cal H}\to \gamma\gamma$ and ${\cal H}\to Z\gamma$ are given by  
\begin{align}
\Gamma({\cal H} \to \gamma \gamma)&=
\frac{\sqrt{2}G_F\alpha_{\text{em}}^2m_{\cal H}^3}{256\pi^3 }\times 
 \Bigg|\xi_{\cal H}F_{\text{SM}}
 -Q_{2+}^2\sum_{\phi=S,\kappa}\frac{\lambda_{{\cal H} \phi^{++}\phi^{--}}}{v}F_0^{\cal H}(m_{\phi^{\pm\pm}})\notag\\
& 
+ \bar{\xi}_{\cal H}Q_{3-}^2 N_E g_S^{}\left(\frac{v}{M_L}\right)F_{1/2}^{\cal H}(M_L)
+ Q_{2-}^2N_E \sum_{\alpha =1,2} y_{{\cal H}E_\alpha E_\alpha}^{} \left(\frac{v}{M_{E_\alpha}}\right) F_{1/2}^{\cal H}(M_{E_\alpha})
 \Bigg|^2, \label{Hgg}   \\
\Gamma({\cal H} \to Z\gamma)&=
\frac{\sqrt{2}G_F\alpha_{\text{em}}^2m_h^3}{128\pi^3}\left(1-\frac{m_Z^2}{m_h^2}\right)^3
\times\Bigg|
\xi_{\cal H}G_{\text{SM}}-(-s_W^2Q_{2+}^2)\sum_{\phi = S,\kappa}\frac{\lambda_{{\cal H}\phi^{++}\phi^{--} }}{v} G_0^{\cal H}(m_{\phi^{\pm\pm}})\notag\\
& +\bar{\xi}_{\cal H} Q_{3-}(-1/2-s_W^2Q_{3-}) N_E g_S^{} \left(\frac{v}{M_L}\right) G_{1/2}^{\cal H}(M_L)\notag\\
& + Q_{2-}\left(\frac{\sin^2\theta_E}{2}-s_W^2Q_{2-}\right)N_E\,  y_{{\cal H}E_1 E_1}^{}\left(\frac{v}{M_{E_1}}\right) G_{1/2}^{\cal H}(M_{E_1})\notag\\
& + Q_{2-}\left(\frac{\cos^2\theta_E}{2}-s_W^2Q_{2-}\right)N_E\,  y_{{\cal H}E_2 E_2}^{} \left(\frac{v}{M_{E_2}}\right) G_{1/2}^{\cal H}(M_{E_2})\notag\\
& + Q_{2-}  \frac{\sin2\theta_E}{4}N_E\,  y_{{\cal H}E_1 E_2}^{} G_{1/2}^{\cal H}(M_{E_1},M_{E_2})     \Bigg|^2, \label{HZg}
\end{align}
where $\bar{\xi}_{\cal H}= \cos\alpha\,(-\sin\alpha)$ for ${\cal H}=H\,(h)$
and $Q_X$ denotes the electric charge, i.e., $Q_t=2/3$, $Q_b=-1/3$, $Q_{3-}=-3$ and $Q_{2\pm}=\pm 2$. 
In the above formulae, 
The Yukawa couplings $y_{{\cal H}E_\alpha E_\beta}^{}$ and the scalar trilinear couplings $\lambda_{{\cal H} \phi^{++}\phi^{--}}$ are given by
\begin{align}
y_{{\cal H}E_1 E_1}^{} &=  \frac{y_E^{}}{\sqrt{2}}\xi_{\cal H} \sin2\theta_E  + g_S^{}\bar{\xi}_{\cal H} ,     \\
y_{{\cal H}E_2 E_2}^{} &=  -\frac{y_E^{}}{\sqrt{2}}\xi_{\cal H} \sin2\theta_E + g_S^{}\bar{\xi}_{\cal H} ,     \\
y_{{\cal H}E_1 E_2}^{} &=   \frac{y_E^{}}{\sqrt{2}}\xi_{\cal H} \cos2\theta_E ,     \\
\lambda_{{\cal H} \kappa^{++}\kappa^{--} } &= -(v\lambda_{\Phi \kappa} \xi_{\cal H} + A_{\Sigma \kappa} \bar{\xi}_{\cal H}) ,  \\
\lambda_{{\cal H} S^{++}S^{--} } &=  -(v\lambda_{\Phi S} \xi_{\cal H} + A_{\Sigma S} \bar{\xi}_{\cal H}). 
\end{align}
The contribution of the SM particles to  ${\cal H} \to \gamma\gamma$ ($F_{\text{SM}}$) and 
${\cal H} \to Z\gamma$ ($G_{\text{SM}}$) are expressed as
\begin{align}
F_{\text{SM}} &= F_1^{\cal H}(m_{W}) +3\sum_{f=t,b} Q_{f}^2 F_{1/2}^{\cal H}(m_f) , \\
G_{\text{SM}} &= G_1^{\cal H}(m_{W}) +3\sum_{f=t,b}Q_f\left(\frac{I_f}{2} -s_W^2Q_f\right)G_{1/2}^{\cal H}(m_f), 
\end{align}
with $I_f = 1/2\,(-1/2)$ for $f=t\,(b)$. 
The loop functions for the $\gamma\gamma$ mode are expressed by  
\begin{align}
F_0^{\cal H}(m_{\varphi})  & = \frac{2v^2}{m_{\cal H}^2}[1+2m_{\phi^\pm}^2C_0(0,0,m_{\cal H}^2,m_{\varphi},m_{\varphi},m_{\varphi})],\\
F_{1/2}^{\cal H}(m_F)  & = -\frac{4m_F^2}{m_{\cal H}^2}\left[2-m_{\cal H}^2\left(1-\frac{4m_F^2}{m_{\cal H}^2}\right)C_0(0,0,m_{\cal H}^2,m_F,m_F,m_F)\right], \\
F_1^{\cal H}({m_W})  & = \frac{2m_W^2}{m_{\cal H}^2}\left[6+\frac{m_{\cal H}^2}{m_W^2}+(12m_V^2-6m_{\cal H}^2)C_0(0,0,m_{\cal H}^2,m_W,m_W,m_W)\right], 
\end{align}
and those for the $Z\gamma$ mode are given by  
\begin{align}
G_0^{\cal H}(m_{\varphi})  &=\frac{2v^2}{e(m_{\cal H}^2-m_Z^2)}
\Bigg\{1+2m_{\varphi}^2C_0(0,m_Z^2,m_{\cal H}^2,m_{\varphi},m_{\varphi},m_{\varphi}) \notag\\
&+\frac{m_Z^2}{m_{\cal H}^2-m_Z^2}\left[B_0(m_{\cal H}^2,m_{\varphi},m_{\varphi})-B_0(m_Z^2,m_{\varphi},m_{\varphi})\right]\Bigg\}, \\
G_{1/2}^{\cal H}(m_{F}) &= \frac{4m_F^2}{s_Wc_W}(4C_{23}+4C_{12}+C_0)(0,m_Z^2,m_{\cal H}^2,m_F,m_F,m_F), \\
G_{1/2}^{\cal H}(m_{F_1},m_{F_2}) &= \frac{4v}{s_Wc_W}\left[2(m_{F_1}+m_{F_2})C_{23}+2(m_{F_1}+m_{F_2})C_{12}+m_{F_1}C_0\right]\notag\\
&\hspace{20mm}(0,m_Z^2,m_{\cal H}^2,m_{F_1},m_{F_2},m_{F_2}) + (F_1 \leftrightarrow F_2), \\
G_1^{\cal H}(m_W^{}) &=\frac{2m_W^2}{s_Wc_W(m_h^2-m_Z^2)}\Bigg\{\left[c_W^2\left(5+\frac{m_{\cal H}^2}{2m_W^2}\right)-s_W^2\left(1+\frac{m_{\cal H}^2}{2m_W^2}\right)\right]\notag\\
&\left[1+\frac{m_Z^2}{m_{\cal H}^2-m_W^2}(B_0(m_{\cal H}^2,m_W,m_W)-B_0(m_Z^2,m_W,m_W))\right]\notag\\
&\hspace{-5mm}+[2m_W^2-6c_W^2(m_{\cal H}^2-m_Z^2)+2s_W^2(m_{\cal H}^2-m_Z^2)]C_0(0,m_Z^2,m_{\cal H}^2,m_{W},m_{W},m_{W})\Bigg\}, 
\end{align}
where $B_i$ and $C_{ij}$ are the two- and three-point Passarino-Veltman functions~\cite{PV}, respectively.  
The notation for these functions is the same as that in Ref.~\cite{PV2}. 
In addition to the above mentioned decay modes, the $H \to hh$ mode is generally allowed. 
However, this mode typically reduces the branching fraction of the $H \to \gamma\gamma$ channel to one order, and it makes difficult to 
explain the observed cross section of the diphoton signature. 
We thus assume that the decay rate of this process is zero by taking the dimensionful $Hhh$ coupling to be zero.  

\begin{figure}[t]
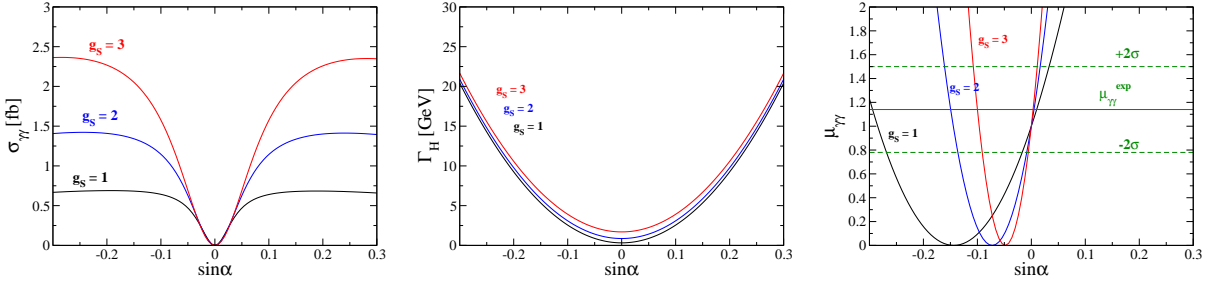

\begin{center}
\includegraphics[scale=0.21]{cross.eps}\hspace{3mm}
\includegraphics[scale=0.21]{width.eps}\hspace{3mm}
\includegraphics[scale=0.21]{strength.eps}
   \caption{The $\sin\alpha$ dependence of 
the cross section $\sigma_{\gamma\gamma}$ for the diphoton process (left), the total width $\Gamma_H$ of $H$ and the signal strength $\mu_{\gamma\gamma}$ (right). 
We take $N_E=3$ and $\lambda_{{\cal H} S^{++}S^{--}}=\lambda_{{\cal H} \kappa^{++}\kappa^{--}}=0$. 
The black, blue and red curves show the case of $g_S^{}=1$, 2 and 3, respectively. 
For the right panel, the central value of and the $2\sigma$ limit on $\mu_{\gamma\gamma}$ from the LHC Run-I experiment are also shown as 
the green horizontal lines. 
 }
   \label{result1}
\end{center}
\end{figure}

Let us perform the numerical analysis to show our predictions of the 
cross section $\sigma_{\gamma\gamma}$ for the diphoton process $gg \to H \to \gamma\gamma$, the total width $\Gamma_H$ of $H$ and the signal strength $\mu_{XY}^{}$. 
In the following analysis, 
we take the mixing angle $\theta_E$ to be zero (equivalently taking $y_E^{}=0$), where a non-zero value of $\theta_E$ does not give an important change of the value of 
$\Gamma({\cal H}\to \gamma\gamma)$ and $\Gamma({\cal H}\to Z\gamma)$. 
We also take all the masses of the exotic leptons and the doubly-charged scalar bosons to be 375 GeV which maximizes the value of $\Gamma(H \to \gamma\gamma)$ for a given 
set of other fixed parameters. 

In Fig.~\ref{result1}, we show the $\sin\alpha$ dependence for the diphoton cross section $\sigma_{\gamma\gamma}$  (left panel), 
the total width $\Gamma_H$ (center panel) and the signal strength $\mu_{\gamma\gamma}$ (right panel) in the case of the number of flavor of the exotic leptons $N_E$ to be 3. 
In these plots, we take $\lambda_{{\cal H} S^{++}S^{--}}=\lambda_{{\cal H} \kappa^{++}\kappa^{--}}=0$, in which 
only the exotic leptons give the additional contributions to the ${\cal H}\to \gamma\gamma$ and ${\cal H}\to Z\gamma$ decays. 
The value of the Yukawa coupling $g_S^{}$ is taken to be 1, 2 and 3 in all the panels. 
For the right panel, 
the measured value of $\mu_{\gamma\gamma}$, i.e., $\mu_{\gamma\gamma}^{\text{exp}}=1.14\pm 0.76$~\cite{Abe} 
at the LHC Run-I experiment is also shown, where the solid and dashed curves denote the central value and the $2\sigma$ limit, respectively. 
We obtain the cross section to be about $0.6$, $1.4$ and $2.4$ fb when $|\sin\alpha|\gtrsim 0.1$, 0.15 and 0.2 in the case of $g_S=1$, 2 and 3, respectively. 
Regarding to the width $\Gamma_H$, its value strongly depends on $\sin\alpha$, while the dependence on $g_S$ is quite weak. 
We find that $\Gamma_H\simeq 2.4\,(8.5)$ GeV at $|\sin\alpha|=0.1\,(0.2)$ with $g_S=1$. 
For $\sigma_{\gamma\gamma}$ and $\Gamma_H$, the sign of $\sin\alpha$ does not become important so much, while that for $\mu_{\gamma\gamma}$ does quite important. 
This can be understood in such a way that the interference effect in the $h \to \gamma\gamma$ process 
between the $W$ boson loop and the exotic lepton loops becomes constructive (destructive) when 
$\sin\alpha$ is positive (negative). 
Because of this destructive effect, the value of $\mu_{\gamma\gamma}$ becomes zero at $\sin\alpha \lesssim 0$, and it rapidly grows when $\sin\alpha$ 
is taken to be a different value from that giving $\mu_{\gamma\gamma}=0$. 
Therefore, the case with $\sin\alpha$ taken to be a bit different value from that giving $\mu_{\gamma\gamma}=0$ is allowed 
by the current experimental data $\mu_{\gamma\gamma}^{\text{exp}}$. 
For the other signal strengths which have been measured at LHC, i.e., $\mu_{ZZ}$, $\mu_{WW}$ and $\mu_{\tau\tau}$, 
they are calculated by $\cos^2\alpha$ at the tree level. 
In the range of $\sin\alpha$ that we take in Fig.~\ref{result1}, we obtain $\cos^2\alpha>0.91$, so that 
these signal strengths are allowed at the $2\sigma$ level from the LHC Run-I data~\cite{LHC_ATLAS2,LHC_CMS2}.

\begin{figure}[t]
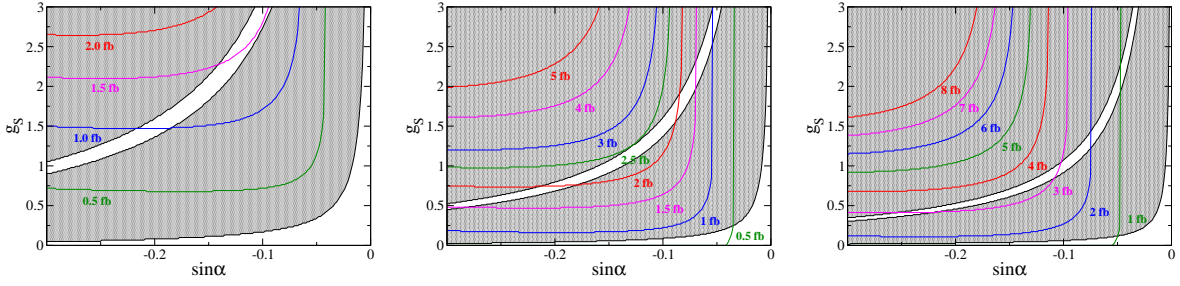

\begin{center}
\includegraphics[scale=0.21]{contour1.eps}\hspace{3mm}
\includegraphics[scale=0.21]{contour2.eps}\hspace{3mm}
\includegraphics[scale=0.21]{contour3.eps}
   \caption{Contour plots for the cross section $\sigma_{\gamma\gamma}$ on the $\sin\alpha$-$g_S^{}$ plane. 
We take $\lambda_{{\cal H} S^{++}S^{--}}=\lambda_{{\cal H} \kappa^{++}\kappa^{--}}=0$. 
The left, center and right panels respectively show the case of $N_E=3$, 6 and 9. 
 }
   \label{result2}
\end{center}
\end{figure}

In Fig.~\ref{result2}, we show the contour plots of $\sigma_{\gamma\gamma}$ on the $\sin\alpha$-$g_S^{}$ plane in the case of 
$\lambda_{{\cal H} S^{++}S^{--}}=\lambda_{{\cal H} \kappa^{++}\kappa^{--}}=0$. 
The left, center and right panels respectively show the case of $N_E=3$, 6 and 9. 
We restrict the range of $\sin\alpha$ to be $0$ to $-0.3$, because the positive value of $\sin\alpha$ is highly disfavored by $\mu_{\gamma\gamma}^{\text{exp}}$ 
as we see in Fig.~\ref{result1}. 
The shaded region is excluded by $\mu_{\gamma\gamma}^{\text{exp}}$ at the $2\sigma$ level. 
We find that the maximally allowed value of the cross section $\sigma_{\gamma\gamma}$ is about 1.5 fb, 2.5 fb and 3 fb when $N_E$ is taken to be 3, 6 and 9, respectively. 
 
\begin{figure}[t]
\begin{center}
\includegraphics[scale=0.35]{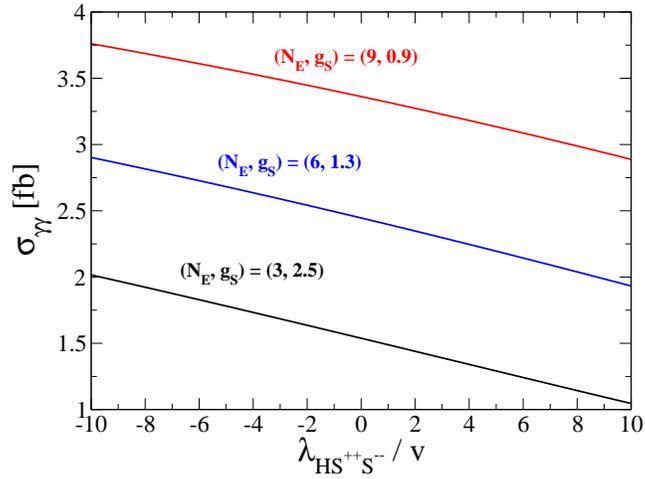}
   \caption{
Cross section $\sigma_{\gamma\gamma}$ as a function of $\lambda_{HS^{++}S^{--}}/v$. 
We take $\lambda_{H\kappa^{++}\kappa^{--}} = \lambda_{HS^{++}S^{--}}$ and $\lambda_{h\kappa^{++}\kappa^{--}} = \lambda_{hS^{++}S^{--}}=0$. 
The black, blue and red curves respectively show the case of $(N_E,g_S^{})=(3,2.5)$, (6,1.3) and (9,0.9). 
For all the plots, we take $\sin\alpha=-0.12$.  }
   \label{result3}
\end{center}
\end{figure}

Finally, we add the non-zero contributions to ${\cal H} \to \gamma\gamma$ from the doubly-charged scalar bosons $S^{\pm\pm}$ and $\kappa^{\pm\pm}$. 
In Fig.~\ref{result3}, we show the diphoton cross section $\sigma_{\gamma\gamma}$ as a function of $\lambda_{H S^{++}S^{--}}(=\lambda_{H \kappa^{++}\kappa^{--}}$
normalized by $v$ 
in the case of $\lambda_{h S^{++}S^{--}}=\lambda_{h \kappa^{++}\kappa^{--}}$. 
In this case, only the $H\to \gamma\gamma /Z\gamma$ mode is modified as compared to the previous cases shown in Figs.~\ref{result1} and \ref{result2} for 
the same parameter choice of $N_E$ and $g_S^{}$. 
In this figure, we take $(N_E,g_S^{})=(3,2.5)$, (6,1.3) and (9,0.9), and $\sin\alpha=-0.12$ for these three cases, where 
these points give the maximal allowed value of $\sigma_{\gamma\gamma}$ that is found in Fig.~\ref{result2}. 
We can see that the constructive effect between the exotic lepton loops and the doubly-charged scalar boson loops 
is obtained when $\lambda_{h S^{++}S^{--}}<0$. 
At $\lambda_{h S^{++}S^{--}}/v = -10$, we obtain $\sigma_{\gamma\gamma}\simeq 2.0$, 2.8 and 3.8 fb at $(N_E,g_S^{})=(3,2.5)$, (6,1.3) and (9,0.9), respectively.

\section{Conclusions}

We have constructed the three-loop neutrino mass model whose structure is similar to the model by Krauss, Nasri and Trodden. 
The neutrino masses of ${\cal O}(0.1)$ eV are naturally generated
by the loop effect of new particles with their couplings and masses to be of order 0.1-1 and TeV, respectively. 
We have analyzed the Majorana DM candidate, assuming the lightest of $N_R$. The non-relativistic cross section to explain the observed relic density is $p$-wave dominant, and there are several processes; $N_{R}^1N_{R}^1  \to \kappa^{++}\kappa^{--}$ with the $t-$ and $u-$channels, 
and $N_{R}^1N_{R}^1 \to f\bar{f}$, $N_{R}^1N_{R}^1 \to ZZ$, $N_{R}^1N_{R}^1 \to W^+W^-$, $N_{R}^1N_{R}^1 \to hh$, $N_{R}^1N_{R}^1 \to HH$ with the $s$-channel. 
The dominant DM scattering with a nucleus comes from the Higgs boson mediation $h$ and $H$ at the tree level, and 
we have calculated the spin independent cross section of the process. 
Furthermore, the anomaly of the muon $g-2$ can be solved by the one-loop contribution of the triply-charged exotic leptons and doubly-charged scalar boson. 
We have found the benchmark parameter set to satisfy the relic abundance of the DM, the constraint from the direct search experiment and 
to compensate the deviation in the measured value of the muon $g-2$ from the SM prediction. 

We then have numerically shown the cross section of the diphoton process via the gluon fusion production $gg \to H \to \gamma\gamma$ and the width of $H$ under 
the constraint from the signal strength $\mu_{\gamma\gamma}$ for the SM-like Higgs boson measured at the LHC Run-I experiment. 
We have obtained the width to be about 3-5 GeV in the typical parameter region, which gives a tension to the measured value, i.e., about 45 GeV. 
We have found that the cross section of the diphoton process is given to be a few fb level 
by taking the masses of new charged fermions and scalar bosons to be 375 GeV with an order 1 coupling constant.  
A bit larger cross section such as about 4 fb is obtained by taking the larger number of flavor $N_E$ of the exotic leptons and take 
a non-zero negative value of the trilinear scalar boson couplings $\lambda_{HS^{++}S^{--}}$ and  $\lambda_{H\kappa^{++}\kappa^{--}}$. 
\\\\
{\it Acknowledgments}

H.O. expresses his sincere gratitude toward all the KIAS members, Korean cordial persons, foods, culture, weather, and all the other things.
K.Y. is supported by JSPS postdoctoral fellowships for research abroad.

\end{document}